\pdfoutput=1
\documentclass[conference]{IEEEtran}
\usepackage{graphicx}

\usepackage[hyphens]{url}
\PassOptionsToPackage{hyphens}{url}\usepackage[hidelinks]{hyperref}
\usepackage{booktabs} 
\usepackage{multirow}
\usepackage{makecell}
\usepackage{listings}
\usepackage{bigstrut}
\usepackage{subcaption}
\usepackage{xspace}
\usepackage[T1]{fontenc}
\usepackage[font=footnotesize,labelfont=bf]{caption}
\usepackage{listings}
\usepackage{amsfonts}
\usepackage{amsmath}
\usepackage{algpseudocode}
\usepackage{algorithm}
\usepackage{numprint}
\npthousandsep{,}

\usepackage[dvipsnames]{xcolor}
\usepackage[utf8]{inputenc}

\usepackage{color}
\definecolor{red}{rgb}{1,0,0}
\definecolor{green}{rgb}{0,1,0}
\definecolor{blue}{rgb}{0,0,1}
\definecolor{cyan}{rgb}{0.4,1,1}
\definecolor{orange}{rgb}{1,0.6,0}
\definecolor{dkgreen}{rgb}{0,0.6,0}
\definecolor{dkred}{rgb}{0.6,0,0}
\definecolor{gray}{rgb}{0.5,0.5,0.5}
\definecolor{purple}{rgb}{0.58,0,0.82}

\lstdefinelanguage{esm}{
    comment=[l]{;},
    commentstyle=\color{gray},
    string=[s]{(}{)},
    stringstyle=\color{gray}
}

\DeclareMathOperator*{\argmin}{arg\,min}

\newif\ifDRAFT
\DRAFTtrue
\DRAFTfalse

\ifDRAFT
  
  \newcommand{\fix}[1]{\textcolor{red}{#1}}
  \newcommand{\todo}[1]{{\color{red}\bf\em TODO:\/\@#1}}
  
  \newcommand{\empirical}[1]{\fbox{#1}}
\else
  \newcommand{\empirical}[1]{#1}
  
  \newcommand{\fix}[1]{{}}
  \newcommand{\todo}[1]{{}}
  
\fi

\newcommand{\Slowdown}{100\xspace}

\newcommand{\point}[1]{\par\smallskip\noindent\textbf{#1.}}
\newcommand{\op}[1]{\lstinline{#1}}

\newcommand{\ETHRate}{145}

\AtEndDocument{\par\leavevmode}

\newcommand{\Months}{2.5\xspace}
 
\begin{document}
\sloppy
\title{Broken Metre:\\ Attacking Resource Metering in EVM}


\author{%
\IEEEauthorblockN{Daniel Perez}
\IEEEauthorblockA{Imperial College London}
\and
\IEEEauthorblockN{Benjamin Livshits}
\IEEEauthorblockA{Imperial College London, \\UCL Centre for Blockchain Technologies, and Brave Software
}}
\IEEEoverridecommandlockouts
\makeatletter\def\@IEEEpubidpullup{6.5\baselineskip}\makeatother
\IEEEpubid{\parbox{\columnwidth}{
    Network and Distributed Systems Security (NDSS) Symposium 2020\\
    23-26 February 2020, San Diego, CA, USA\\
    ISBN 1-891562-61-4\\
    https://dx.doi.org/10.14722/ndss.2020.24267\\
    www.ndss-symposium.org
}
\hspace{\columnsep}\makebox[\columnwidth]{}}

\maketitle



\begin{abstract}
  Blockchain systems, such as Ethereum, use an approach called ``metering'' to assign a cost to smart contract execution, an approach which is designed to incentivise miners to operate the network and protect it against DoS attacks. In the past, the imperfections of Ethereum metering allowed several DoS attacks which were countered through modification of the metering mechanism.

  This paper presents a new DoS attack on Ethereum which systematically exploits its metering mechanism. We first replay and analyse several months of transactions, during which we discover a number of discrepancies in the metering model, such as significant inconsistencies in the pricing of the instructions. We further demonstrate that there is very little correlation between the execution cost and the utilised resources, such as CPU and memory. Based on these observations, we present a new type of DoS attack we call~\emph{Resource Exhaustion Attack}, which uses these imperfections to generate low-throughput contracts. To do this, we design a genetic algorithm that generates contracts with a throughput on average~\Slowdown times slower than typical contracts. We then show that all major Ethereum client implementations are vulnerable and, if running on commodity hardware, would be unable to stay in sync with the network when under attack. We argue that such an attack could be financially attractive not only for Ethereum competitors and speculators, but also for Ethereum miners. Finally, we discuss short-term and potential long-term fixes against such attacks. Our attack has been responsibly disclosed to the Ethereum Foundation and awarded a bug bounty reward of~5,000~USD.
\end{abstract}

%


\pagestyle{plain} 
\section{Introduction}
\label{sec:introduction}

Some blockchain systems support code execution, allowing arbitrary programs to take advantage of decentralised trust. Ethereum and its virtual machine, the Ethereum Virtual Machine (EVM), is probably the most widely  used blockchain adopting this approach. However, allowing arbitrary programs from non-trusted users introduces many new challenges. One of these challenges is to prevent users from running code which could negatively impact the performance of the system. To tackle this challenge, Ethereum introduced the notion of ``gas'', which is a unit used to measure the execution cost of a program, referred to as a ``smart contract'' in this context. Gas-based metering is used to price the execution of smart contracts, and must ensure that the throughput of the blockchain, in terms of gas per second, remains stable. Metering is therefore critical to keep the Ethereum blockchain safe against Denial of Service (DoS) attacks involving slow running contracts. However, assigning costs to different instructions is a highly non-trivial task, and the costs originally assigned in the Ethereum yellow paper~\cite{wood2014ethereum}, which were designed to maintain a throughput of 1 gas/$\mu$s, had many inconsistencies. As a consequence, several DoS attacks have been conducted on Ethereum~\cite{transaction-spam-attack,suicide-attack}, and the gas cost has also been reviewed several times~\cite{erc150,eip-1884} to increase the cost of the under-priced instructions.

To the best of our knowledge, there has still not been any attempt to try to find and exploit such inconsistencies in a systematic way. In this paper, we design a new DoS attack which exploits inconsistencies of the gas metering mechanism by taking a systematic approach to finding these. We first replay and analyse several months of transactions to discover discrepancies in the gas cost. We then use the data and insight from our analysis to design a genetic algorithm capable of generating low-throughput contracts. We evaluate the contracts generated by our algorithm on all major Ethereum clients and find that they are all vulnerable to our attack.

\point{Contributions}
This paper makes the following contributions:
\begin{enumerate}
    \item \textbf{Exploration of metering in EVM}: We explore the history of executing~\empirical{\Months} months worth of smart contracts on the Ethereum blockchain and identify several important edge cases that highlight inherent flaws in EVM metering; specifically, we identify~i) EVM instructions for which the gas fee is too low compared to their resources consumption; and~ii) cases of programs where the cache influences execution time by an order of magnitude.
    
	
    \item \textbf{Resource Exhaustion Attacks (REA) contract generation strategy}: We present a code generation strategy able to produce REA attacks of arbitrary length. Some of the complexity comes from the need to produce well formed EVM programs which minimise the throughput. We propose an approach which combines empirical data and a genetic algorithm in order to generate contracts with low throughput. We explore the efficacy of our strategy as a function of the throughput in terms of gas per second of the generated programs.
	\item \textbf{Experimental evaluation}:
	We show that our REA can abuse imperfections in EVM's metering approach. Our genetic algorithm is able to generate programs with a throughput of~\empirical{1.25M} gas per second after a single generation. A minimum in our experiments is attained at generation~\empirical{243} with a block using around~\empirical{9.9M} gas and taking about~\empirical{93 seconds}. We show that our method generates contracts on average more than~\Slowdown times slower than typical contracts. Finally, we evaluate our low-throughput contracts on the major Ethereum clients and show that they are all vulnerable. Using commodity hardware, nodes would be unable to stay in sync when under attack.
\item \textbf{Disclosure and fixes}:
  We responsibly disclosed our attack to the Ethereum foundation, and were awarded a bug bounty reward of~5,000~USD. We discussed with the developers about the ongoing efforts as well as some potential fixes, and present some of the short-term and long-term fixes in this paper.
\end{enumerate}

\point{Paper Organisation} The rest of the paper is organised as follows. In Section~\ref{sec:background}, we provide background information about Ethereum and its metering scheme, as well as a few instances of how it has been exploited in the past. In Section~\ref{sec:case-studies}, we present case studies based on measurements that we obtained by re-executing the Ethereum main chain. In Section~\ref{sec:attack}, we present our Resource Exhaustion Attacks (REA) and the results we obtained. In Section~\ref{sec:design} we present short and long-term solutions to gas mispricing issues. Finally, we present related work in Section~\ref{sec:related} and conclude in Section~\ref{sec:conclusion}.

\section{Background}
\label{sec:background}
In this section, we briefly describe the Ethereum network and the EVM. Then, we provide an in-depth explanation of how the gas mechanism works and provide additional insights into smart contract execution costs on the Ethereum main network. Finally, we highlight some of the attacks which have been performed by abusing the gas mechanism.

\subsection{Ethereum and the Ethereum Virtual Machine (EVM)}
The Ethereum~\cite{Buterin2014} platform allows its users to run ``smart contracts'' on its distributed infrastructure. Ethereum \emph{smart contracts} are programs which define a set of rules for the governing of associated funds, typically written in a Turing-complete programming language called Solidity~\cite{Dannen:2017:IES:3103305}. The Solidity code is compiled into EVM bytecode, a low level bytecode designed to be executed by the EVM.

Once the EVM bytecode is generated, it is deployed on the Ethereum blockchain by sending a transaction which only purpose is to create a smart contract with the given code. To execute a smart contract, a user can then send a transaction to this contract. The sender will pay a \emph{transaction fee} which is derived from the contract's computational cost, measured in units of~\emph{gas}~\cite{wood2014ethereum}. The fee itself is paid in Ether (ETH\footnote{When converting ETH to USD, we use the exchange rate on 2020-01-07: 1 ETH = \ETHRate{} USD. For consistency, any monetary amounts denominated in USD are based on this rate.}), the underlying currency of the Ethereum blockchain. When a \emph{miner} successfully mines a blocks, he receives the transaction fee of all the transactions included in the block. We will describe exactly how this transaction fee is computed in the following part of this section.

\subsection{Metering in EVM}
As briefly outlined in Section~\ref{sec:introduction}, gas is a fundamental component of Ethereum, and generally applicable to permissioned and permissionless blockchain platforms that utilise a distributed virtual machine for contract code execution~\cite{tezos-about,eosio-about}. Gas is the main protection against Denial of Service (DoS) attacks based on non-terminating or resource-intensive programs. It is also used to incentivise miners to process transactions by rewarding them with a fee computed based on the resource usage of the transaction.

\point{Gas cost}
In the EVM, each transaction has a cost which is computed in and expressed as gas. The cost is split into two parts, a fixed \textit{base cost} of $21,000$ gas, and a variable \emph{execution cost} of the smart contract. 
Each instruction has a fixed gas cost which has been set by the designers of the EVM~\cite{wood2014ethereum}, who classify the instructions in multiple tiers of gas cost: zero Tier (0 gas), base tier (2 gas), very low tier (3 gas), low tier (5 gas), high tier (10 gas) and special tier where the cost needs more complex rules.
\begin{figure}[tb]
  \begin{center}
    \begin{minipage}{0.42\textwidth}
\begin{lstlisting}[language=esm]
PUSH1 0x02 ; very low tier (3 gas)
PUSH1 0x03 ; very low tier (3 gas)
MUL        ; low tier      (5 gas)
PUSH1 0x05 ; very low tier (3 gas)
SSTORE     ; special tier  (20k gas)
\end{lstlisting}
\end{minipage}
\caption{Example gas cost of an EVM program}
\label{list:example-gas-cost}
\end{center}
\end{figure}
The gas cost for a transaction in the EVM is the sum over the cost of each instruction in the contract. For example, given the program in Figure~\ref{list:example-gas-cost}, the gas cost will be computed as follow. \op{PUSH1} is in the Very Low Tier and therefore costs~3 gas. It is called~3 times in total and will therefore consume~9 gas. The arguments of \op{PUSH1} do not consume any extra gas. The \op{MUL} instruction is in the Low Tier and hence costs 5 gas. Finally, the \op{SSTORE} will store the result of~$2\times3$ at location 5 in the storage. \op{SSTORE} is in the Special Tier and has slightly more complex pricing rules. Assuming the location in the storage was previously~0, the instruction allocates storage and will cost~20,000 gas. Therefore, this program will cost a total of~20,014 gas to execute. Given the current pricing for storage, the cost of the program is clearly dominated by the storage operation.

It is important to note that, as the transaction has a base cost of~21,000 gas, it will cost a total of~$21\text{,}000 + 20\text{,}014 = 41\text{,}014$ gas to execute the above transaction.

\point{Ethereum Improvement Proposal~(EIP)~150}
Although the cost of each instruction was decided when first designing the EVM, the authors found that some costs were poorly aligned with actual resource consumption. Particularly, IO-heavy instructions tended to be too cheap, allowing for DOS attacks on the Ethereum~\cite{suicide-attack} blockchain. As a fix, EIP~150~\cite{erc150} was proposed and implemented, significantly increasing the gas consumption of instructions which require to perform IO operations, such as \lstinline{SLOAD} or \lstinline{EXTCODESIZE}. This change revised the cost of under-priced instructions and prevented further DoS attacks such as the one seen in September~2016~\cite{transaction-spam-attack}. This briefly highlights the potential risks rooted in mismatches between instructions and gas costs. While the above cases have been fixed, it is unclear whether all potential issues have been eradicated or not.

\point{Gas price} Up to here, we have explained how the gas cost for executing a contract are computed. However, the gas cost is not the only element needed to compute the total execution cost of a contract. When a transaction is sent, the sender can choose a gas price, namely the amount of \emph{wei} ($1\text{wei} = 10^{-18}~\text{ETH}$) that the sender is ready to pay per unit of gas. For conciseness, these amounts are often expressed in Gwei, where $1\text{Gwei} = 10^9\text{wei}$. Miners will usually prioritise transactions with high gas prices, as this will increase the final fee they receive for processing a transaction.

\point{Transaction fee}
The transaction fee is the total amount of wei that the sender of the transaction has to pay for the transaction. It is obtained by multiplying the gas price by the gas cost. The transaction fee is non-refundable: even if the transaction fails, it will be paid.

\begin{figure}[tb]
  \centering
  \setlength{\tabcolsep}{10pt}
  \begin{tabular}{lrr}
    \toprule
     & \multicolumn{2}{c}{Gas price}\\
     & Low & High\\
    Transaction type & (1Gwei) & (80Gwei)\\
    \midrule
    Basic (21k gas) & \$0.00 & \$0.24\\
    Gas intensive (500k gas) & \$0.07 & \$5.80\\
    \bottomrule
  \end{tabular}
  \caption{Fees for different type of transactions. ``Low'' price is one of the lowest possible price to have a transaction included while ``High'' is a price someone very eager to have his transaction included would pay.}
  \label{tab:gas-fee}
\end{figure}
\begin{figure}[tb]
\setlength{\tabcolsep}{3pt}
\centering
\begin{tabular}{lr}
    \toprule
    Number of blocks: & 613,475\\
    Median gas price: & 9.1 Gwei\\
    Median gas used (by contracts): & 53,787 \\
    Median transaction fee: &  0.0008 ETH (0.12~USD)\\
    \bottomrule
\end{tabular}
\caption{Median gas price, gas used and transaction fee from block 8,652,096 (Sep-09-2019) to block 9,286,594 (Jan-15-2020).}
\label{tab:empirical-gas-fee}
\end{figure}

\subsection{Gas Statistics}
Now that we presented the key points about metering in the EVM, we provide concrete numbers about different aspects of the gas price and transaction fees. In particular, we show the amount of transaction fees that a user would have to pay to have his transaction processed by the main Ethereum network.

To give a sense of the transaction fees, we show a variety of typical fees in Figure~\ref{tab:gas-fee}. The fees are divided depending on their gas price and gas consumption. The \textit{Low} gas price is close to the lowest price that can be paid to get the transaction accepted on the Ethereum blockchain. The \textit{High} gas price refers to the price that people would pay when they are extremely eager to get their transaction included, for example when competing with other users to have a transaction included first~\cite{gas-price-history}. The \textit{basic} transaction type refers to transactions consuming only the base amount of gas, without executing any instruction. This is typically the cost to send Ether to a contract or another party. The \textit{gas intensive} transaction type represents computationally expensive transactions, for example, verifying a zero-knowledge proof~\cite{aztec-protocol}. At the time of writing, the maximum amount of gas which can be used in a single block is~\empirical{10,000,000}, which means only~\empirical{20} such transactions could be included in a single block.

In Figure~\ref{tab:empirical-gas-fee}, we show the values of the gas price, gas used and transaction fee. In order to obtain results reflecting the current situation, we limit the analysis to recent blocks. We use all the transactions sent to contracts between September 30, 2019 and January 15, 2020. We find that the median gas price paid by a transaction's sender is around~\empirical{9.1} Gwei, which is around~\empirical{9} times more than the minimum possible fee. It is worth noting that when paying the minimum possible fee, the probability for the transaction to get included in the next block is relatively low and the transaction can therefore be delayed for several blocks: at the time of writing, about \empirical{40\%} of the last 200 blocks accepted a gas price of \empirical{1Gwei}~\cite{eth-gas-station}. This explains that users usually pay a higher fee to get their transaction included faster. The median for the gas consumed by contracts is around~\empirical{50,000} gas, indicating that most transactions perform relatively simple computations. Indeed, the basic fee being~21,000, a simple read followed by an allocation of storage would already result in~46,000 gas. Overall, the median fee paid per transactions is~\empirical{0.0008 ETH} which is around~\empirical{0.12 USD}.


\subsection{Previously Known Attacks}
The Ethereum network has been victim of several Denial of Service (DoS) attacks due to instructions being under-priced. We present two considerable DoS attacks which were performed on the Ethereum network.

\point{\lstinline{EXTCODESIZE} attack}
In September 2016, a DoS attack was performed on the Ethereum network by flooding it with transactions containing a very large number of \lstinline{EXTCODESIZE} instructions~\cite{transaction-spam-attack}. \lstinline{EXTCODESIZE} is an instruction to retrieve the size in bytes of a given contract's code.

\begin{figure}[tb]
  \centering\includegraphics[width=\columnwidth]{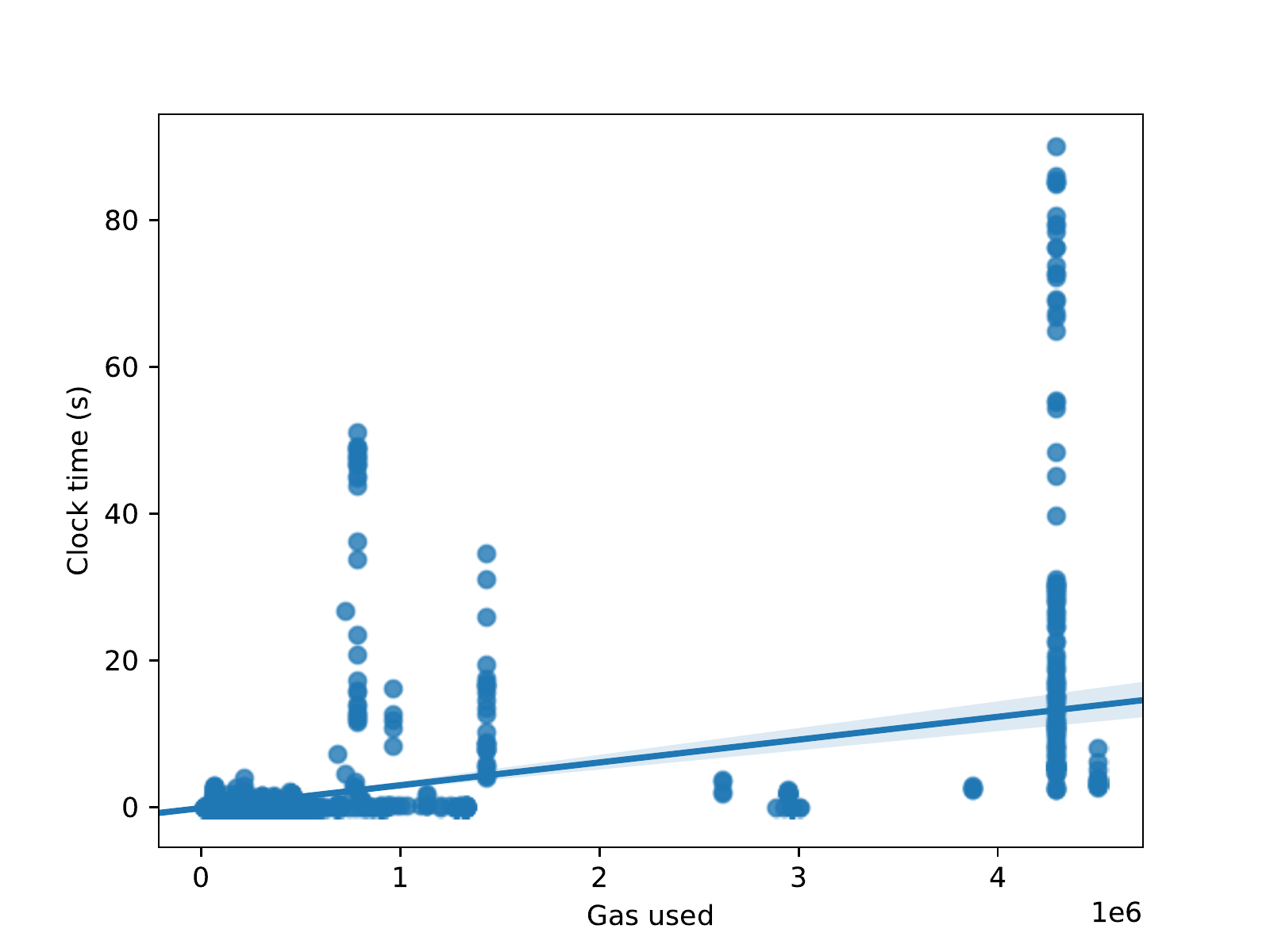}
  \caption{Correlation between gas and clock time with DoS.}
  \label{fig:extcodesize-cpu}
\end{figure}

This attack happened because the \lstinline{EXTCODESIZE} instruction was vastly under-priced. At the time of the attack, a single execution of this instruction cost~20 gas, meaning that one could perform around~1,500 instructions with less than~\$0.01. Although by itself, this issue might seem benign, \lstinline{EXTCODESIZE} forces the client to search the contract on disk, resulting in IO heavy transactions. While replaying the Ethereum history on our hardware, the malicious transactions took around~20 to~80 seconds to execute, compared to a few milliseconds for the average transactions. We show the correlation between the clock time and the gas used by transactions during the period of the attack in Figure~\ref{fig:extcodesize-cpu}. Although this attack did not create any issue at the consensus layer, it reduced the rate of block creation by a factor of more than 2 times, with block creation time peaking to more than~30s~\cite{block-time-chart}.

The Ethereum protocol was updated in EIP~150, with all the software running Ethereum, to increase the price of the \lstinline{EXTCODESIZE} from~20 to~700 gas, making the aforementioned attack considerably more expensive to perform. Some performance improvements were also made at the implementation level, allowing clients to process IO-intensive instructions faster.

\point{\lstinline{SUICIDE} Attack}
Shortly after the \lstinline{EXTCODESIZE} attack, another DoS attack involving the \lstinline{SUICIDE} instruction was performed~\cite{suicide-attack}. The \lstinline{SUICIDE} instruction kills a contract and sends all its remaining Ether to a given address. If this particular address does not exist, a new address would be newly created to receive the funds. Furthermore, at the time of the attack, calling \lstinline{SUICIDE} did not cost any Ether. Given these two properties, an attacker could create and destroy a contract in the same transaction, creating a new contract each time at an extremely low fee. This quickly overused the memory of the nodes, and particularly affected the Go implementation~\cite{geth} which was less memory efficient~\cite{geth-memory-efficiency}.

A twofold fix was issued for this attack in EIP~150. First, and most importantly, \lstinline{SUICIDE} would be charged the regular amount of gas for contract creation when it tried to send Ether to a non-existing address. Subsequently, the price of the \lstinline{SUICIDE} instruction was increased from~0 to~5,000 gas. Again, these measures would make such an attack very expensive.

\section{Case Studies in Metering}
\label{sec:case-studies}

In this section, we instrument the C\texttt{++} client of the Ethereum blockchain, called \textit{aleth}~\cite{aleth}, and report some interesting observations about gas dynamics in practice.

\subsection{Experimental setup}
\point{Hardware} We run all of the experiments on a Google Cloud Platform (GCP)~\cite{gcp-compute-engine} instance with~4 cores~(8 threads) Intel Xeon at~2.20GHz,~8~GB of RAM and an SSD with a~400MB/s throughput. The machine runs Ubuntu 18.04 with the Linux kernel version~4.15.0. We selected this hardware because it is representative to what has been reported as sufficient to run a full Ethereum node~\cite{node-incentive,pantheon-system-requirements,eth-hardware-requirements}.

\point{Software} To measure the speed of different instructions, we fork the Ethereum C\texttt{++} client, \textit{aleth}. Our fork integrates the changes to the upstream repository until Jun-26 2019. We choose the C\texttt{++} client for two reasons: first, it is one of the two clients officially maintained by the Ethereum Foundation~\cite{ethereum-foundation-github} with geth~\cite{geth}; second, it is the only of the two without runtime or garbage collection, which makes measuring metrics such as memory usage more reliable.

We add compile options to the original C\texttt{++} client to allow enabling particular measurements such as CPU or memory. Our measurement framework is open-sourced\footnote{\url{https://github.com/danhper/aleth/tree/measure-gas}} and available under the same license as the rest of \textit{aleth}.

\point{Measurements}
For all our measurements, we only take into account the execution of the smart contracts and ignore the time taken in networking or other parts of the software. We use a nanosecond precision clock to measure time and measure both the time taken to execute a single smart contract and the time to execute a single instruction. To measure the memory usage of a single transaction, we override globally the \lstinline[language=C++]{new} and \lstinline[language=C++]{delete} operators and record all allocations and deallocations performed by the EVM execution within each transaction. We ensure that this is the only way used by the EVM to perform memory allocation.

Given the relatively large amount of time it takes to re-execute the blockchain, we only execute each measurement once when re-executing. We ensure that we always have enough data points, where enough in the order of millions or more, so that some occasional imprecision in the measurements, which are inevitable in such experiments, do not skew the data.

In this section, the measurements are run between block~\empirical{5,171,468 (Feb-28-2018)} and block~\empirical{5,587,480 (May-10-2018)}, except in~\ref{ssec:system-resources} where we want to compare after and before EIP-150.

\begin{figure}[tb]
  \begin{subfigure}[t]{\columnwidth}
    \centering\includegraphics[width=\columnwidth]{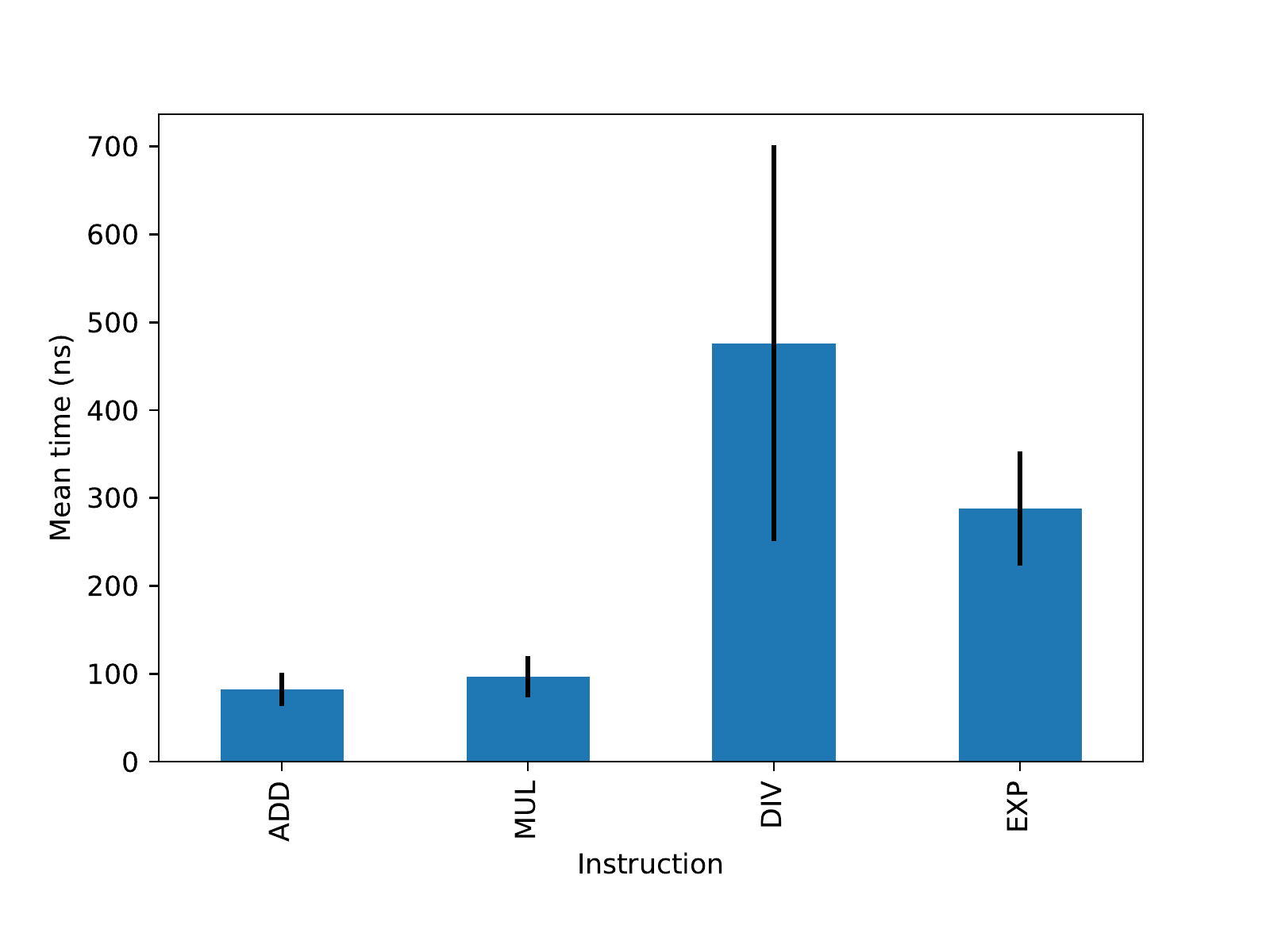}
    \vskip -5mm
    \caption{Mean time for arithmetic instructions.}
    \vskip 5mm
    \label{fig:arithmetic-instruction-times}
  \end{subfigure}
  \begin{subfigure}[t]{\columnwidth}
    \centering
    \begin{tabular}{crrrr}
      \toprule
      \bf \multirow{2}{*}{Instruction} & \bf Gas & \bf \multirow{2}{*}{Count} & \multicolumn{1}{c}{\bf Mean} & \bf Throughput \\
      & \bf cost & & \textbf{time} (ns) &(gas / $\mu$s)\\
      \midrule
      \texttt{ADD} & 3  & 453,069 & 82.20 & 36.50\\
      \texttt{MUL} & 5  & 62,818 & 96.96 & 51.57\\
      \texttt{DIV} & 5  & 107,972 & 476.23 & 10.50\\
      \texttt{EXP} & \textasciitilde 51 & 186,004 & 287.93 & 177.1\\
      \bottomrule
    \end{tabular}
    \caption{Execution time and gas usage for arithmetic instructions.}
    \label{tab:arithmetic-instruction-times}
  \end{subfigure}
  \caption{Comparing execution time and gas usage of arithmetic instructions.}
\end{figure}

\subsection{Arithmetic Instructions}
In this experiment, we evaluate the correlation between gas cost and the execution time for simple instructions which include absolutely no IO access. We use simple arithmetic instructions for measurements, in particular the \lstinline{ADD}, \lstinline{MUL}, \lstinline{DIV} and \lstinline{EXP} instructions.

In Figure~\ref{fig:arithmetic-instruction-times}, we show the mean time of execution for these instructions, including the standard deviation for each measurement. We contrast these results with the gas cost of the different instructions in Figure~\ref{tab:arithmetic-instruction-times}. \lstinline{EXP} is the only of these instructions with a variable cost depending on its arguments --- the value of the exponent. We use the average gas cost in our measurements to compute the throughput. We see that although in practice \lstinline{ADD} and \lstinline{MUL} have similar execution time, the gas cost of \lstinline{MUL} is~65\% higher than the gas cost for \lstinline{ADD}. On the other hand, \lstinline{DIV}, which costs the same amount of gas as \lstinline{MUL}, is around~\emph{5 times slower} on average. \lstinline{EXP} costs on average \emph{10 times} the price of \lstinline{DIV} but executes~40\% faster. Another point to note here is that \lstinline{DIV} has a standard deviation much higher than the other three instructions. Although we were expecting that for such simple instructions the execution time would reflect the gas cost, this does not appear to be the case in practice. We will show in the coming sections that IO related operations tend to make things worse in this regard.


\begin{figure}[tb]
  \centering
  \setlength{\tabcolsep}{14pt}
  \begin{tabular}{clr}
    \toprule
    \thead[l]{Phase} & \thead[l]{Resource} & \thead[r]{Pearson\\score}\\
    \midrule
    \multirow{4}{*}{Pre EIP-150} & Memory & 0.545\\
    & CPU & 0.528\\
    & Storage & 0.775\\
    & Storage/Memory & \textbf{0.845}\\
    & Storage/Memory/CPU & 0.759\\
    \midrule
    \multirow{4}{*}{Post EIP-150} & Memory & 0.755\\
    & CPU & 0.507\\
    & Storage & 0.907\\
    & Storage/Memory & \textbf{0.938}\\
    & Storage/Memory/CPU & 0.893\\
    \bottomrule
  \end{tabular}
  \caption{Correlation scores between gas and system resources.}
  \label{tab:correlation-scores}
\end{figure}

\subsection{Gas and System Resources Consumption}
\label{ssec:system-resources}
In this section, we analyse the gas consumption of Ethereum smart contracts and try to correlate it with different system resources, such as memory, CPU and storage.
As described in Section~\ref{sec:background}, EIP-150 influenced the price of many storage related operations, which affected the gas cost of transactions. Therefore, we use a different set of transactions than for other case studies. We arbitrarily use block~\empirical{1,400,000} to block ~\empirical{1,500,000} for measurements before EIP-150 and block ~\empirical{2,500,000} to~\empirical{2,600,000} for measurements after EIP-150. We assume that the sample of~\empirical{100,000} blocks, which roughly corresponds to two weeks, is large enough to obtain reliable data.

We use our modified Ethereum client to perform the different measurements. To measure memory, we compute the difference between the total amount of memory allocated and the total amount of memory deallocated. For CPU, we use clock time measurements as a proxy for the CPU usage. Finally, for storage usage, we count the number of EVM words~(256 bits) of storage newly allocated per transactions.

We compute the Pearson correlation coefficient\footnote{Pearson score of 1 means perfect positive correlation, 0 means no correlation}~\cite{boslaugh2012statistics} between the different resources and the gas usage. We also compute multi-variate correlation between gas consumption and multiple resources. To compute the multi-variate correlation between multiple resources and the gas usage, we first normalise the measurement vector of each targeted resource to have a mean of $0$ and a standard deviation of $1$. Then, we stack the vectors to obtain a matrix of $m$ resources and $n$ measurements, and transform it in a single vector of $n$ measurements using a principal component analysis~\cite{abdi2010principal}. The vector we obtain represents the aggregated usage of the different resources and can be correlated with the gas usage.

We present our results in~\autoref{tab:correlation-scores}. A first observation is that EIP-150 clearly emphasises the domination of storage in the price of contracts. We can clearly see that storage alone has an extremely high correlation score, with score of~\empirical{0.907} after EIP-150. Memory usage is not as correlated as storage, but when combining both, they have the highest correlation score of~\empirical{0.938}. Finally, an important point is that CPU time seems completely uncorrelated with gas usage. Although it seems natural that CPU time by itself has a low correlation, as gas cost is dominated by storage cost, adding the CPU time in the multi-variate correlation reduces the correlation. It is not enough to make any conclusion yet but gives a hint that as long as the storage is not explicitly touched, it could be possible for contracts to be both cheap and long to execute.

\subsection{High-Variance Instructions in the EVM}
Here, we look at instructions which have a high variance in their execution time. We summarise the instructions which had the highest variance in Figure~\ref{tab:high-variance-instructions}. There are two main reasons why the execution time may vastly vary for the execution of the same instruction. First, many instructions take parameters, depending on which, the time it takes to run the particular instructions can vary wildly. This is the case for an instruction such as \lstinline{EXTCODECOPY}. The second reason is much more problematic and comes from the fact that some instructions may require to perform some IO access, which can be influenced by many different factors such as caching, either at the OS or at the application level. The instruction with the highest variance was \lstinline{BLOCKHASH}. 
\lstinline{BLOCKHASH} allows to retrieve the hash of a block and allows to look up to~256 block before the current one. When it does so, depending on the implementation and the state of the cache, the EVM may need to perform an IO access when executing this instruction, which can result in vastly different execution times. The cost of \lstinline{BLOCKHASH} being currently fixed and relatively cheap,~20 gas, it results in an instruction which is vastly under-priced. It is worth noting that in the particular case of \lstinline{BLOCKHASH}, the issue has already been raised more than two years ago in EIP-210~\cite{eip-blockhash}. It discussed changing the price of \lstinline{BLOCKHASH} to~800 gas but at the time of writing the proposal is still in draft status and was not included in the Constantinople fork\footnote{Hard fork which took place on Feb 28 2019 on the Ethereum main network}~\cite{constantinople} as it was originally planned to be.

\begin{figure}[tb]
  \centering
    \setlength{\tabcolsep}{3pt}
  \begin{tabular}{lrrr}
    \toprule
    \textbf{Instruction} & \textbf{Mean} & \textbf{Standard} & \textbf{Measurements}\\
    & \textbf{time} ($\mu$s) & \textbf{deviation} & \textbf{count}\\
    \midrule
    \texttt{BLOCKHASH} & 768 & 578 & 240,000\\
    \texttt{BALANCE} & 762 & 449 & 8,625,000\\
    \texttt{SLOAD} & 514 & 402 & 148,687,000\\
    \texttt{EXTCODECOPY} & 403 & 361 & 23,000\\
    \texttt{EXTCODESIZE} & 221 & 245 & 16,834,000\\
    \bottomrule
    \end{tabular}
    \caption{Instructions with the highest execution time variance.}
    \label{tab:high-variance-instructions}
\end{figure}


\begin{figure}[tb]
  \includegraphics[width=\columnwidth]{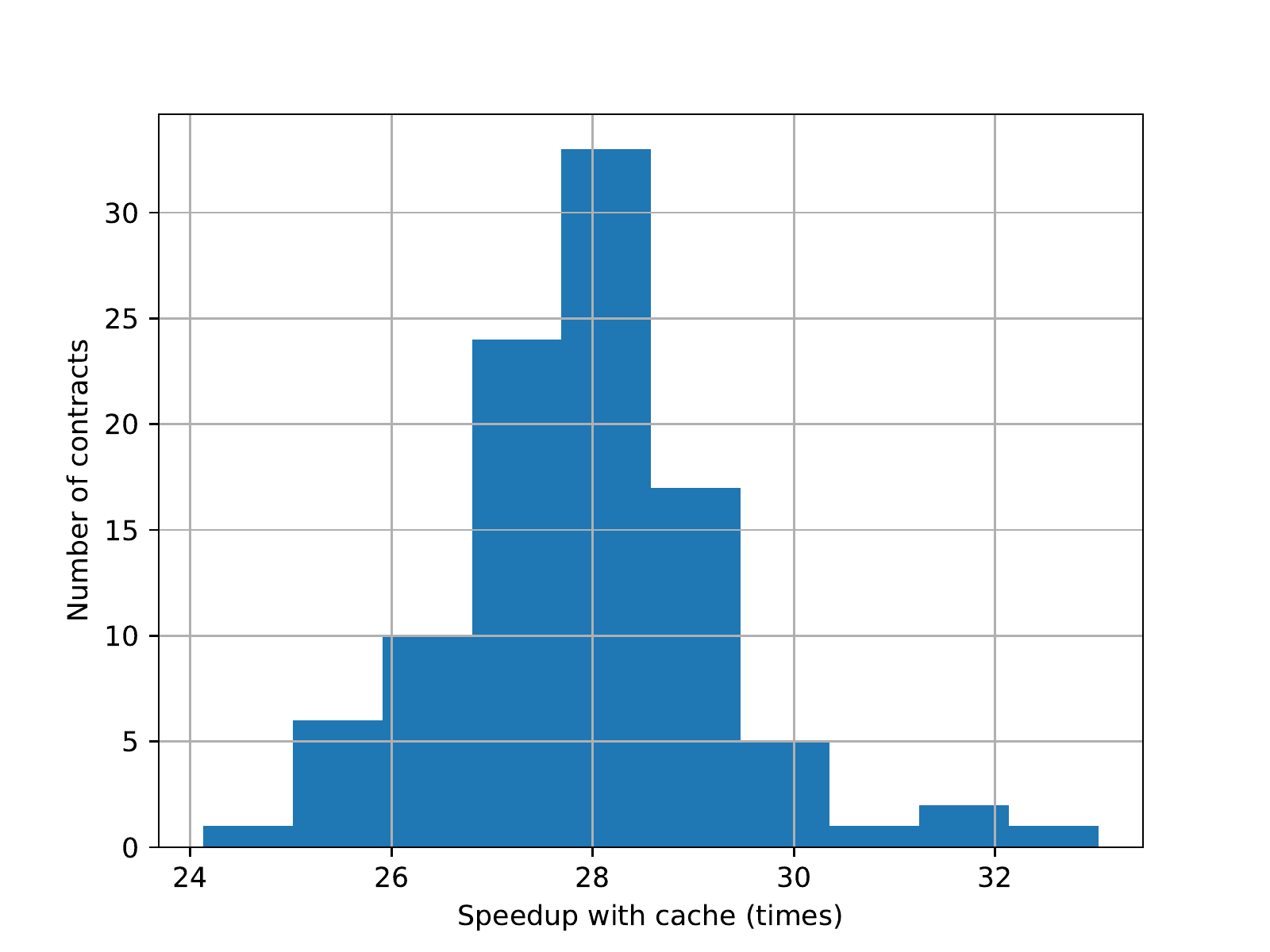}
  \caption{Comparing throughput with and without page cache: $x$ axis is the relative speed improvement and $y$ axis is the number of contracts.}
  \label{fig:cache-measurement}
\end{figure}

\begin{figure}[tb]
  \includegraphics[width=\columnwidth]{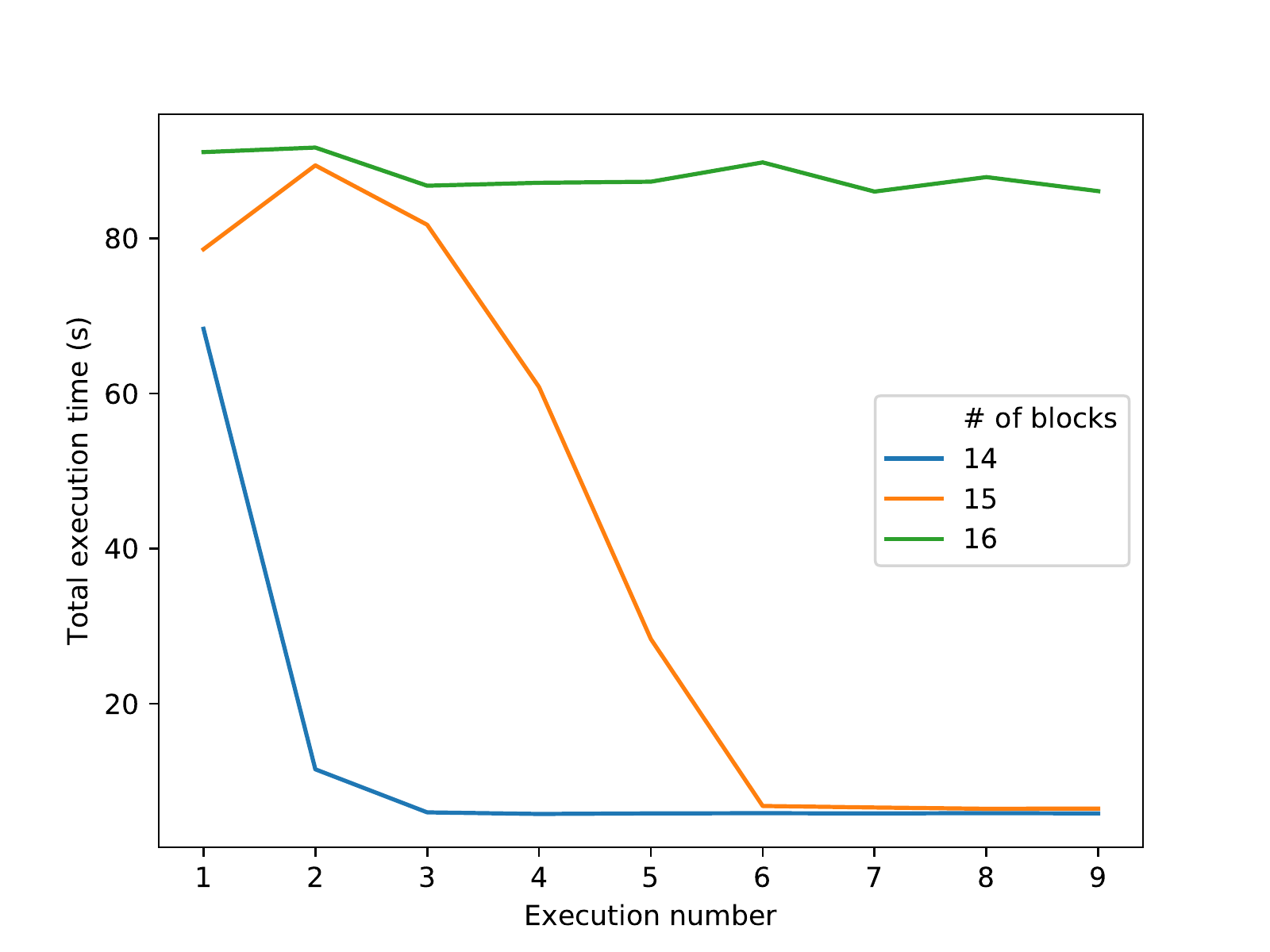}
  \caption{Measuring block execution speed with and without the effect of cache.}
  \label{fig:cache-persistence}
\end{figure}

\subsection{Memory Caches and EVM Costs}
\label{ssec:memory-caches}
Given the high variance in execution time for some instructions, we evaluate the effects caching may have on EVM execution speed. In particular, we evaluate both the speedup provided by the operating system page cache and the speedup across blocks provided by LevelDB LRU cache~\cite{leveldb-cache}. In these experiments, we fix the block number at height~\empirical{5,587,480}.

\point{Page cache} First, we evaluate how the operating system page cache influences the execution time by reducing the IO latency. We perform the experiment as follows:

\begin{enumerate}
\item Generate a contract
\item Run the code of the contract $n$ times
\item Run the code of the contract $n$ times but drop the page cache between each run
\end{enumerate}
We perform this for~\empirical{100} different contracts and measure the execution time for the versions with and without cache. We generate relatively large contracts, which consume on average~\empirical{800,000} gas each. Although the method is somewhat crude, it provides a good approximation of the extent to which the state of the page cache influences the execution time of a contract. In Figure~\ref{fig:cache-measurement}, we show a distribution of the contracts throughput in terms of gas per second, with and without cache. We see that contracts execute between~\empirical{24} and~\empirical{33} times faster when using the page cache, with more than half of the contracts executing between~\empirical{27} and~\empirical{29} times faster.
This vast difference in the execution speed is due to IO operations, which use LevelDB~\cite{ghemawat2011leveldb}, a key-value store database, under the hood. LevelDB keeps only a small part of its data in memory and therefore needs to perform a disk access when the data was not found in memory. If the required part of the data was already in the page cache, no disk access will be required. When keeping the page cache, all the items seen by the contract recently will already be available in cache, eliminating the need for any disk access. On the other hand, if the caches are dropped, many IO related operations will result in a disk access, which explains the speedup. We notice that in the contracts with the highest speedup, \lstinline{BLOCKHASH}, \lstinline{BALANCE} and \lstinline{SLOAD} are in the most frequent instructions. It is worth noting that if the generated contracts are small enough, most of the data will be in memory and dropping the page cache will have much less effect on the runtime. Indeed, when running the same experiment with contracts consuming on average~\empirical{100,000} gas, only a 2 times average speedup has been observed.
\point{Caching across blocks} In the next experiment, instead of measuring the cache impact by running a single contract multiple times, we evaluate how the cache impacts the execution time across blocks. In particular, we measure how many blocks need to be executed before the data cached during the previous execution of a contract gets evicted from the different caches. To do so, we perform the following experiment.

\begin{enumerate}
\item Generate $n$ blocks, with different contracts in each
\item Execute sequentially all the blocks and measure the execution time
\item Repeat the previous step $m$ times in the same process, and record how the execution speed evolves
\end{enumerate}

We set $m$ to 10 and we try different values for $n$ to see how many blocks are needed for the cache not to provide anymore speedup. We use the first execution to warm-up the node and use the $9$ other executions for our measurements. We find that in our setup, assuming the blocks are full (i.e. close to the gas limit in term of gas), $16$ blocks are enough for the cache not to provide anymore speedup. We plot the results for \empirical{$n = 14$}, \empirical{$n = 15$} and \empirical{$n = 16$} in Figure~\ref{fig:cache-persistence}. When \empirical{$n = 14$}, we see that the second execution is much faster than the first one, and that after the third execution, the execution time stabilises at around \empirical{6s} to execute the \empirical{$14$} blocks. For \empirical{$n = 15$}, the execution time takes longer to decrease, but eventually also stabilises around the same value. It is slightly higher than when \empirical{$n = 14$} because it has one more block to execute. However, once we reach \empirical{$n = 16$}, we see that the execution time hardly decreases and stays stable at around \empirical{85s}. We conclude that at this point, almost nothing that was cached during the previous execution of the block is still cached when re-executing the block.

This means that if a deployed contract function were re-executed more than 16 blocks after its initial execution, it would execute as slowly as the first time.
This shows that not only the cache has a very high impact on execution time but also that the cached information is evicted relatively quickly.

\subsection{Summary}
In this section, we empirically analysed the gas cost and resource consumption of different instructions. To summarise:
\begin{itemize}
\item We see that even for simple instructions, the gas cost is not always consistent with resource usage. Indeed, even for instruction with very predictable speed, such as arithmetic operations, we observe that some instructions have a throughput~\empirical{5} times slower than others.
\item We notice that while most instructions have a relatively consistent execution speed, other instructions have large variations in the time it takes to execute. We find that these instructions involve some sort of IO operation.
\item Finally, we demonstrate the effect that the page cache has on the execution speed of smart contracts and then show the typical number of blocks for which the page cache still provides speed up.
\item Overall, we see that beyond some pricing issue, the metering scheme used by EVM does not allow to express the complexity inherent to IO operations.
\end{itemize}

\section{Attacking the Metering Model of EVM}
\label{sec:attack}
In light of the results we obtained in the previous sections, we hypothesise that it is possible to construct contracts which use a low amount of gas compared to the resources they use. 

\subsection{Constructing Resource Exhaustion Attacks}
In particular, as we showed in Section~\ref{sec:case-studies}, the gas consumption is dominated by the storage allocated but is not as much affected by other resources such as the clock time. Therefore, we decide to use the clock time as a target resource and look for contracts which minimise the throughput in terms of gas per second. We can formulate this as a search problem.

\point{Problem formulation}
We want to find a program which has the minimum possible throughput, where we define the throughput to be the amount of gas processed per second.
Let $\mathbb{I}$ be the set of EVM instructions and $P$ be the set of EVM programs. A program $p\in P$ is a sequence of instructions $I_1,\cdots,I_n$ where all $I_i \in \mathbb{I}$. Let $t : P \rightarrow \mathbb{R}$ be a function which takes a program as an input and outputs its execution time and $g : P \rightarrow \mathbb{N}$ be a function which takes a program as input and outputs its gas cost. We define our function to minimise $f: P \rightarrow \mathbb{R},~f(p) = g(p) / t(p)$. Our goal is to find the program $p_{\text{slowest}}$ such that

\begin{equation}
  \label{eq:objective}
  p_{\text{slowest}} = \argmin_{p\in P} (f(p))
\end{equation}

The search space for a program of size $n$ is $|\mathbb{I}|^n$. Given $|\mathbb{I}| \approx 100$, the search space is clearly too large to be explored entirely for any non-trivial program. Therefore, we cannot simply go over the space of possible programs and instead need to approximate the solution.

Although our problem resembles other program synthesis tasks~\cite{gulwani2017program}, there is a notable difference. Program synthesis usually focuses on generating ``meaningful'' programs, either from specifications or examples. These tasks often do not have well-defined metrics allowing optimisation techniques (the genetic algorithm in our work). The task we solve is different because we need to define ``valid'' but not ``meaningful'' programs and optimise for a well-defined metric: gas throughput.

\point{Search strategy}
With the problem formulated as a search problem, we now present our search strategy. We decide to design the search as a genetic algorithm~\cite{whitley1994genetic}. The reasons for this choice are as follow:

\begin{itemize}
\item we have a well-defined fitness function $f$
\item we have promising initialisation values, which we will discuss below
\item programs being a sequence of instructions, cross-over and mutations can be designed efficiently
\item generated programs need to be syntactically correct but do not need to be semantically meaningful, making the cross-over and mutations more straightforward to design
\end{itemize}
We will now discuss in detail how we design the initialisation, cross-over and mutations of our genetic algorithm.

\point{Program construction}
Although our programs do not need to be semantically meaningful, they need to be executed successfully on the EVM, which means that they must fulfil some conditions. First, an instruction should never try to consume more values than the current number of elements on the stack. Second, instructions should not try to access random parts of the EVM memory, otherwise the program could run out of gas straight away. Indeed, when an instruction reads or writes to a place in memory, the memory is ``allocated'' up to this position and a fee is taken for each allocated memory word. This means that if \lstinline{MLOAD} would be called with $2^{100}$ as an argument, it would result in the cost of allocating $2^{100}$ words in memory, which would result in an out of gas exception.

Another potential issue would be to run into an infinite loop. However, we decide to explicitly exclude loops out of our program generation algorithm for the following reason: adding loops is unlikely to make the generated programs slower. Indeed, if a piece of code is slow enough, our genetic algorithm will tend to repeat it. The loop version could be faster if a value is already cached but have no reason to be slower.

We design the program construction so that all created programs will never fail because of either of the previous reasons. We first want to ensure that there are always enough elements on the stack to be able to execute an instruction. The instructions requiring the least number of elements on the stack are instructions such as \lstinline{PUSH} or \lstinline{BALANCE} which do not require any element, and the element requiring the most number of elements on the stack is \lstinline{SWAP16} which requires $17$ elements to be on the stack. We define the functions function $a : \mathbb{I} \rightarrow \mathbb{N}$ which returns the number of arguments consumed from the stack and $r : \mathbb{I} \rightarrow \mathbb{N}$ which returns the number of elements returned on the stack for an instruction $I$. We generate 18 sets of instructions using Equation~\ref{eq:instr-args}.

\begin{equation}
  \label{eq:instr-args}
  \forall n \in [0, 17],~ \mathbb{I}_n = \{I~|~I\in \mathbb{I} \land a(I) \leq n\}
\end{equation}

For example, the set $\mathbb{I}_3$ is composed of all the instructions which require $3$ or less items on the stack. We will use these sets in Algorithm~\ref{alg:construct-program} to construct the initial programs but before, we need to define the functions we use to control memory access. For this purpose, we define two functions to handle these. First, $uses\_memory : \mathbb{I} \rightarrow \{0, 1\}$ returns $1$ only if the given instruction accesses memory in some way. Then, $prepare\_stack: \mathbb{P}\times\mathbb{I}\rightarrow \mathbb{P}$ takes a program and an instruction and ensures that all the arguments of the instruction which influence which part of memory is accessed are below a relatively low value, that we arbitrarily set to $255$. To ensure this, $prepare\_stack$ adds \lstinline{POP} instruction for all arguments of $I$ and add the exact same number of \lstinline{PUSH1} instructions with a random value below the desired value. For example, in the case of \lstinline{MLOAD}, a \lstinline{POP} followed by a \lstinline{PUSH1} would be generated.

Using the sets $\mathbb{I}_n$, the $uses\_memory$ and $prepare\_stack$ functions, we use Algorithm~\ref{alg:construct-program} to generate the program. We assume that the $biased\_sample$ function returns a random element from the given set and will discuss how we instantiate it in the next section.

\begin{algorithm}
  \begin{algorithmic}
    \Function{GenerateProgram}{$size$}
    \State $P \gets (~)$\Comment{Initial empty program}
    \State $s\gets 0$\Comment{Stack size}
    \For{$1$ to $size$}
      \State $I \gets biased\_sample(\mathbb{I}_s$)
      \If{$uses\_memory(I)$}
        \State $P \gets prepare\_stack(P, I)$
      \EndIf
      \State $P \gets P \cdot (~I~)$\Comment{Append $I$ to $P$}
      \State $s \gets s + (r(I) - a(I))$
    \EndFor
    \State \textbf{return} $P$
    \EndFunction
  \end{algorithmic}
  \caption{Initial program construction}
  \label{alg:construct-program}
\end{algorithm}

\point{Initialisation}
As the search space is very large, it is important to start with good initial values so that the genetic algorithm can search for promising solutions. For this purpose, we use the result of the results we presented in Section~\ref{sec:case-studies}, in particular, we use the throughput measured for each instruction. We define a function $throughput : \mathbb{I} \rightarrow \mathbb{R}$ which returns the measured throughput of a single instruction. When randomly choosing the instructions with $biased\_sample$, we want to have a higher probability of picking an instruction with a low throughput but want to keep a high enough probability of picking a higher throughput instruction to make sure that these are not all discarded before the search begins. We define the weight of each instruction and then its probability with equations~\ref{eq:initial-weight} and~\ref{eq:initial-prob}.

\begin{align}
  \label{eq:initial-weight}
  W(I\in \mathbb{I}) &= \log\left(1 + \frac{1}{throughput(I)}\right)\\
  \label{eq:initial-prob}
  P(I\in \mathbb{I}_n) &= \frac{W(I)}{\sum_{I'\in \mathbb{I}_n}W(I')}
\end{align}
Given that the throughput can have order-of-magnitude differences among instructions, the $\log$ in Equation~\ref{eq:initial-weight} is used to avoid assigning a probability to close to~$0$ to an instruction.

\point{Cross-over}
We now want to define a cross-over function over our search-space, a function which takes as input two programs and returns two programs, i.e. $cross\_over : \mathbb{P} \times \mathbb{P} \rightarrow \mathbb{P} \times \mathbb{P}$, where the output programs are combined from the input programs. To avoid enlarging the search space with invalid programs, we want to perform cross-over such that the two output programs are valid by construction. As during program creation, we must ensure that each instruction of the output program will always have enough elements on the stack and that it will not try to read or write at random memory locations.

For the memory issue, we simply avoid modifying the program before an instruction manipulating memory or one of the \lstinline{POP} or \lstinline{PUSH1} added in the program construction phase. For the second issue, we make sure to always split the two programs at positions where they have the same number of elements on the stack.

We show how we perform the cross-over in Algorithm~\ref{alg:cross-over}. In the \textproc{CreateStackSizeIndex} function, we create a mapping from a stack size to a set of program counters where the stack has this size. In the \textproc{CrossOver} function, we first create this mapping for both programs and randomly choose a stack size to split the program. We then randomly choose a location from each program with the selected stack size. Note that here, $sample$ assigns the same probability to all elements in the set. Finally, we split each program in two at the chosen position, and cross the programs together.

\begin{algorithm}
  \begin{algorithmic}
    \Function{CreateStackSizeMapping}{$P$}
      \State $S \gets $ \text{empty mapping}
      \State $pc \gets 0$
      \State $s \gets 0$
      \For{$I~\text{in}~P$}
        \If{$s\notin S$}
          \State $S[s] \gets \{\}$
        \EndIf
        \State $S[s] \gets S[s] \cup \{pc\}$
        \State $s \gets s + (r(I) - a(I))$
        \State $pc \gets pc + 1$
      \EndFor
      \State \textbf{return} $S$
    \EndFunction

    \Function{CrossOver}{$P_1, P_2$}
      \State $S_1 \gets$~\Call{CreateStackSizeMapping}{$P_1$}
      \State $S_2 \gets$~\Call{CreateStackSizeMapping}{$P_2$}
      \State $S \gets S_1 \cap S_2$\Comment{Intersection on keys}
      \State $s \gets sample(S)$
      \State $i_1 \gets sample(S_1[s])$
      \State $i_2 \gets sample(S_2[s])$
      \State $P_{11}, P_{12} \gets split\_at(P_1, i_1)$
      \State $P_{21}, P_{22} \gets split\_at(P_2, i_2)$
      \State $P_1' \gets P_{11}\cdot P_{22}$\Comment{Concatenate}
      \State $P_2' \gets P_{21}\cdot P_{12}$
      \State \textbf{return} $P_1',~P_2'$
    \EndFunction
  \end{algorithmic}
  \caption{Cross-over function}
  \label{alg:cross-over}
\end{algorithm}

\point{Mutation}
We use a straightforward mutation operator. We randomly choose an instruction $I$ in the program,  where $I$ is not one of the \lstinline{POP} or \lstinline{PUSH1} instructions added to handle memory issues previously discussed. We generate a set $M_I$ of replacement candidate instructions defined as follow.
\begin{equation}
  \label{ref:mutation-set}
  M_I = \{ I'~|~I'\in \mathbb{I}_{a(I)}\land r(I') = r(I) \}
\end{equation}

In other words, the replacement must require at most the same number of elements on the stack and put back the same number as the replaced instruction. Then, we replace the instruction $I$ by $I'$, which we randomly sample from $M_I$. If $I$ had \lstinline{POP} or \lstinline{PUSH1} associated with it to control memory, we remove them from the program. Finally, if $I^\prime$ uses memory, we add the necessary instructions before it.

\subsection{Effectiveness of Attacks with Synthetic Contracts}
\begin{figure}[tb]
    \centering
    \includegraphics[width=.8\columnwidth]{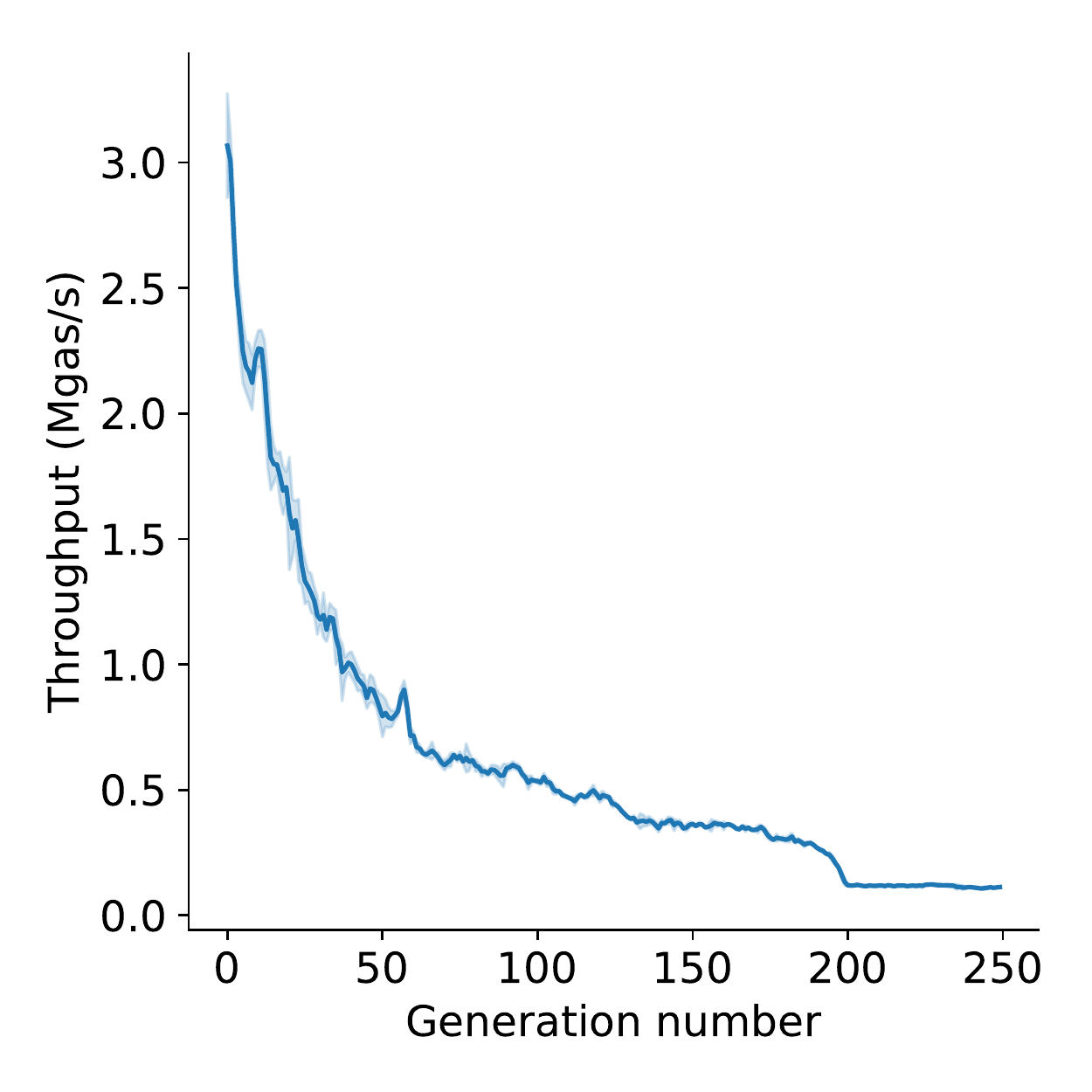}
    \caption{Evolution of the average contract throughput as a function of the number of generations.}
    \label{fig:throughput-evolution}
\end{figure}

\begin{figure}[tb]
    \centering
    \includegraphics[width=.7\columnwidth]{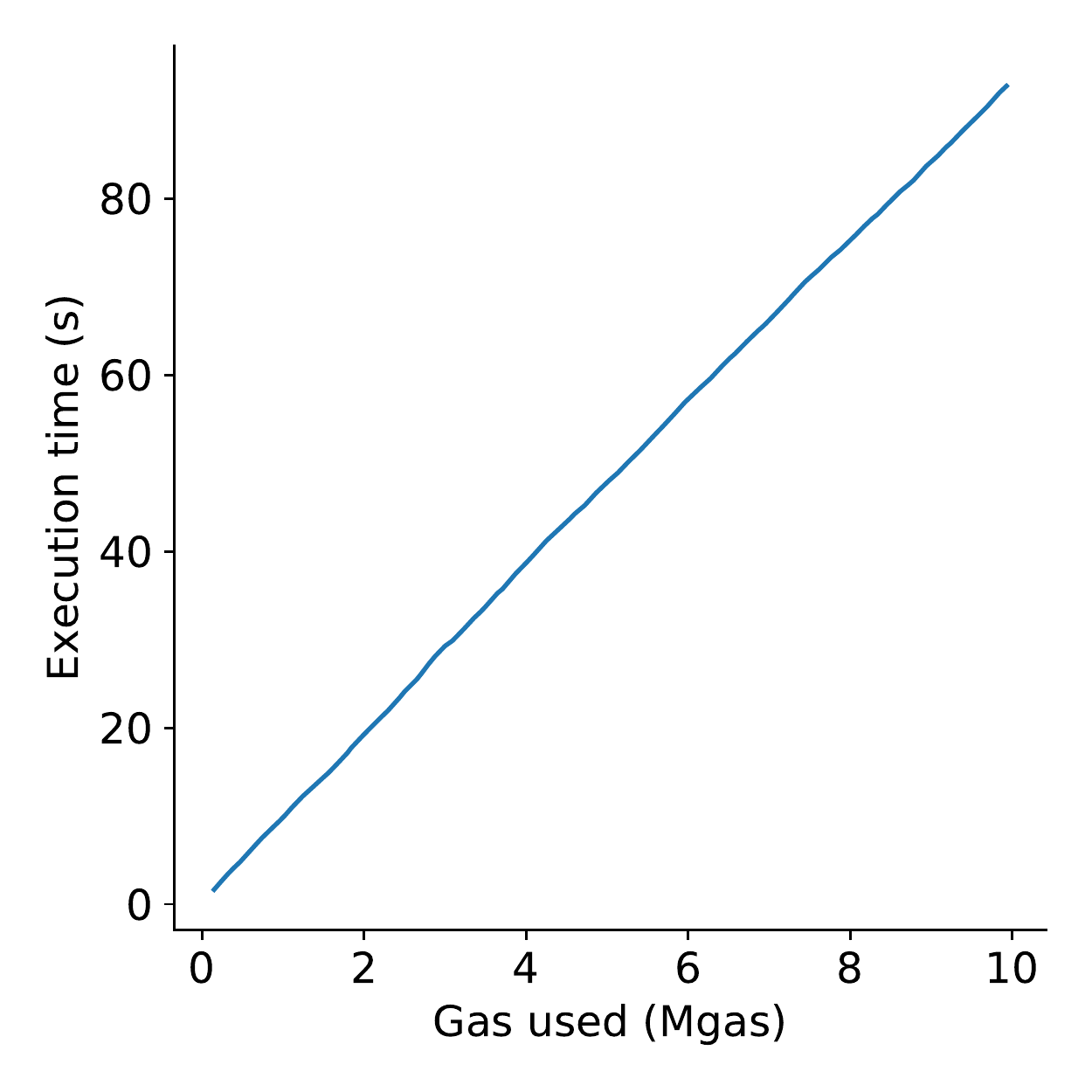}
    \caption{Execution time as a function of amount of gas used by contracts within a block.}
    \label{fig:block-exec-speed}
\end{figure}

We want to measure the effectiveness of our approach to produce Resource Exhaustion Attacks. To do so, we want to generate contracts and benchmark them while mimicking the behaviour of a regular full validating node as much as possible. To do so, we execute all the programs produced within every generation of our genetic algorithm, as if they were part of a single block. We use the following steps to run our genetic algorithm.

\begin{enumerate}
\item Clear the page cache;
\item Warm up caches by generating and executing randomly-generated contracts
\item Generate the initial set of program;
\item Run the genetic algorithm for~$n$ generation.
\end{enumerate}
An important point here is that when running the genetic algorithm, we only want to execute each program once, otherwise every IO access will already be cached and it will invalidate the results, as this is not what would happen when a regular validating node executes contracts. However, we of course do want to execute the measurements multiple times to be able to measure the execution time standard deviation. To work around these two requirements, we save all the programs generated while we run the experiment. Once the experiment has finished, we re-run all the programs in the exact same order. We combine these results to compute the mean and standard deviation of the execution time.

We note that generating a new generation takes on average less than~\empirical{1} second but the time-consuming part of our algorithm is to compute the throughput of the generated programs. Indeed, we need to wait for the EVM to run the program, which can, as we show see in this section, take more than~\empirical{90} seconds for a single generation. Furthermore, parallelising this task could bias our measurements, which forces the algorithm to perform the evaluation serially.

\begin{figure}
\begin{lstlisting}[escapeinside={<@}{@>},language=esm]
<@\footnotesize{PUSH9 0x57c2b11309b96b4c59}@>
<@\textbf{BLOCKHASH}@>
<@\textbf{SLOAD}@>
CALLDATALOAD
<@\footnotesize{PUSH7 0x25dfb360fa775a}@>
<@\textbf{BALANCE}@>
MSTORE8
<@\footnotesize{PUSH10 0x49f8c33edeea6ac2fe8a}@>
<@\footnotesize{PUSH14 0x1d18e6ece8b0cdbea6eb485ab61a}@>
<@\textbf{BALANCE}@>
<@\small{POP}@>      ; prepare call to CALLDATACOPY
<@\small{POP}@>
<@\small{POP}@>
<@\small{PUSH1 0xf7}@>
<@\small{PUSH1 0xf7}@>
<@\small{PUSH1 0xf7}@>
<@\textbf{CALLDATACOPY}@>
<@\footnotesize{PUSH7 0x421437ba67fe0e}@>
<@\footnotesize{ADDRESS}@>
<@\textbf{BLOCKHASH}@>
\end{lstlisting}
  \caption{Bytecode snippet generated by our genetic algorithm. Instructions in bold involve some sort of IO operations.}
  \label{list:generated-code}
\end{figure}

\point{Generated bytecode}
Before discussing the results further, we show a small snippet of bytecode generated by our genetic algorithm in Figure~\ref{list:generated-code}. We highlight the instructions which involve IO operations in bold and show the instructions whose sole purpose is to keep the stack consistent in a smaller font. We can see that there is a large number of IO related instructions, in particular \lstinline{BLOCKHASH} and \lstinline{BALANCE} show up multiple times. Although the fee of \lstinline{BALANCE} has been revised from~20 to~400 in EIP-150, this suggests that the instruction is still under-priced. In the snippet, we also see that the stack is cleared and replaced with small values before calling \lstinline{CALLDATACOPY}. This corresponds to the $prepare\_stack$ function described in the program construction section: to avoid \lstinline{CALLDATACOPY} to read very far away in memory, which would make the program run out of gas, the arguments are replaced with small values. We note that our algorithm can generate programs of arbitrary length but in our experiments we set it to create programs of around~\empirical{4,000} instructions which consume between~\empirical{100,000} and \empirical{150,000} gas.

\point{Generating low-throughput contracts}
We show how the throughput of the lowest performing contract evolved with the number of generations in Figure~\ref{fig:throughput-evolution}. The line represents the mean of the measurements and the band represents the standard deviation of the measurements. The measurements are run~\empirical{3} times. Except from one point in the first measurements, overall the standard deviation remains relatively low. 

We can see that during the first generations, the throughput is around~\empirical{1.25M} gas per second, which is already fairly low given that the average throughput for a transaction on the same machine is around~\empirical{20M} gas per second. This shows that our initialisation is effective. The throughput decreases very quickly in the first few generations, and then steadily decreases down to around~\empirical{110K} gas per second, which is more than~\empirical{180} slower than the average transaction. After about~\empirical{200} generations, the throughput more or less plateaus.  

\point{Exploring the minimum}
The minimum in our experiments is attained at generation~\empirical{243}. At this point, the block uses in total approximately~\empirical{9.9M} gas and takes around~\empirical{93~seconds} to execute, or a throughput of about~\empirical{107,000} gas per second. We show in Figure~\ref{fig:block-exec-speed} how the execution time increases with the amount of gas consumed within the block. It is important to note that the execution time increases perfectly linearly with the gas used, which means that all transactions in the block have almost the same throughput. This implies that an attacker could easily tune the time he wants to delay the nodes depending on his budget. If a block full of malicious transactions were to be processed, given that an Ethereum block is produced roughly every~13 seconds,~\empirical{7} new blocks would have been created by the time the node finishes to validate the malicious one. 

\begin{figure}
  \centering
  \setlength{\tabcolsep}{3pt}
  \begin{tabular}{l r r r}
    \toprule
    \multirow{2}{*}{\textbf{Client}} & \textbf{Throughput} & \textbf{Time} & \textbf{IO load}\\
    & Gas/s & second & MB/s\\
    \midrule
    Aleth        & $107,349\pm 606.6$ & $93.6\pm 0.53$ & $9.12\pm 4.70$\\
    Parity       & $210,746\pm 7,672$ & $47.1\pm 1.61$ & $10.0\pm 1.36$\\
    Parity (\small{metal}) & $542,702\pm 9,487$ & $18.2\pm 0.23$ & $17.2\pm 1.97$\\
    Geth         & $131,053\pm 4,207$ & $75.6\pm 2.42$ & $6.57\pm 4.13$\\
    Geth (\small{fixed})  & $3,021,038 \pm 4.67\mathrm{e}5$ & $3.33 \pm 0.56$ & $0.72\pm 0.11$\\
    \bottomrule
  \end{tabular}
  \caption{Evaluation of different clients when executing 10M gas worth of malicious transactions. What is presented is the mean across \empirical{three} measurements~$\pm$ standard deviation. All the measurements are performed on our GCP except the ``metal'' which is done on our bare-metal server.}
  \label{tab:clients-evaluation}
\end{figure}

\subsection{Evaluation on Other Ethereum Clients}
We used aleth~\cite{aleth} to run our genetic algorithm and find low-throughput contracts. In this section, we show that the contracts crafted using our algorithm are also effective on the two most popular Ethereum clients: geth~(v1.9.6)~\cite{geth} and Parity Ethereum~(v2.5.9)~\cite{parity-ethereum}. We also show that the fix released in geth following discussions with the development team successfully resolves the issue. Furthermore, although our attack is mainly efficient on less powerful hardware, we include the measurements of Parity on a more powerful bare-metal machine with 4 cores (8 threads) at 2.7GHZ, 32GB of RAM and an SSD with 540MB/s throughput.
To benchmark the clients, we use the following procedure, and repeat the measurements \empirical{three} times for each client.

\begin{enumerate}
\item Synchronise the client to test;
\item Start the client in a private network, so that it does not execute anything else but our contracts;
\item Execute transactions on the client using the \lstinline{eth_call} RPC endpoint;
  \begin{enumerate}
  \item Send transactions to warm-up the client
  \item Send enough malicious transactions to consume 10M gas
  \end{enumerate}
\item Measure the gas, time, IO, CPU and memory used during the execution of the malicious transactions.
\end{enumerate}
We report our results in~\autoref{tab:clients-evaluation}. Although we measured CPU, memory, and IO usage, most of the used time was related to IO operations and there was no significant increase in either CPU or memory usage during the attack. Therefore, we only report the IO measurements collected during the attack. We express the IO load in terms of MB/s, which we collect using Linux's \texttt{iotop} utility.

Before geth's fix, geth takes more than~$75$ seconds to execute~10M gas worth of malicious transactions. Parity Ethereum is the least vulnerable to our attack, but still takes on average about~$47$ seconds. Parity has on average a higher, but more constant IO load than geth. Large increase in the IO load tend to increase the IO wait time, which could explain why geth is vastly slower than Parity. Aleth is the slowest of the three clients. There could be two reasons for this: first, our algorithm is optimised on aleth, which makes it more likely to slow it down, second, aleth is less actively developed than the other two clients and might lack some optimisations.

The results of running Parity on a more powerful bare-metal server show that even such machines are relatively vulnerable to our attack. Indeed, Parity, which was the fastest of the tree clients, still took more than~\empirical{18} seconds to execute the transactions. An important point to notice is that the IO throughput is considerably higher on our bare-metal server, which is most likely one of the main reason for the speedup.

Finally, we ran our attack on an improved version of geth, which the Ethereum developers pointed us to as a result of our interactions with them. This version includes several optimisations to improve the storage access speed. We can see that these improvements drastically reduced the IO load of the client. With these improvements, geth executes the transactions more than~20 times faster, making the execution speed fast enough to counter such an attack. Our interaction with geth developers shows the effectiveness of responsible vulnerability disclosure, as discussed in Section~\ref{sec:responsible}. 

\subsection{REA as a Form of DoS}


Malicious contracts crafted using our algorithm could easily be used to perform a DoS attacks on Ethereum. In this section, we will describe the threat model of such an attack, including the implications and feasibility of the attack.

\point{Attack implications} 
As described in Section~\ref{sec:background}, there have already been several instances of DoS attacks against Ethereum~\cite{transaction-spam-attack,suicide-attack}. There are several consequences to such attacks. The most direct one is a high increase in the block production time~\cite{average-block-time}, which in the worst cases more than doubled, significantly decreasing the total throughput of the network. This decrease comes not only from miners who might take more time to validate blocks but also from full nodes who are supposed to relay validated blocks and might take vastly longer to do so. A further indirect consequence of such attacks is the loss of trust in the system, which can lead to a decrease in the price of Ethereum, at least for a short period of time~\cite{Chen2017Metering}.

\point{Probable attacker} 
Although instances of irrational behaviours with likely no profit to the attacker have been seen on Ethereum~\cite{Breidenbach}, we assume that the attacker is rational and wants to profit from such an attack. In this context, there are several ways in which such an attack could be performed.

First, this attack could be beneficial to miners. A miner could use these malicious transactions to perform a sort of selfish-mining~\cite{eyal2014majority}. Indeed, if the miner chooses to include a small amount of malicious transactions in the blocks he mines, the propagation time per block is likely to increase and give the miner a head-start on mining the next block. Given that the block arrival time in Ethereum is around~13 seconds, gaining a couple of seconds can be financially interesting for a miner. Furthermore, the only cost for a miner would be the opportunity cost of not including other transactions in the block, as he could include malicious transactions with a gas price of 0.

Another potential motivation for an attack could be to try to reduce the price of the ETH token and the trust in the Ethereum ecosystem. An attacker wanting to make a one-shot profit could spend some amount of money into performing such a DoS attack while taking a short position on ETH, waiting for the price to go down. Other blockchains competing with Ethereum could also potentially use such tactics to try to discredit the reliability of Ethereum.

\point{Attack feasibility}
To reason about the feasibility of this attack, we assume that given the same gas price, a malicious transaction has the same chance of being included in a block as any other transaction. We use the time we obtained in our experiments with geth, as it is the Ethereum client with the largest usage share~\cite{ehternodes}.

To find a reasonable gas price, we analyse the gas price of all transactions and blocks from~\empirical{October 1, 2019 (block 8,653,171)} to \empirical{December 31, 2019 (block 9,193,265)}. We find that the median value of the minimum gas price in a block is around~\empirical{1.1Gwei} and that the average gas price is around~\empirical{10Gwei} with a standard deviation of~\empirical{11Gwei}. These values are in agreement with some other source of gas computation~\cite{eth-gas-station}. Finally, we find that at least \empirical{2 million} gas worth of transactions are included for less than~\empirical{3Gwei} in about~\empirical{90\%} of the blocks, and choose this value as the gas price to compute the cost of an attack.

Given that our malicious transactions have a throughput of about~131,000 gas per second, using a price of~\empirical{3 Gwei}, it would cost roughly~$131,000 \times 3\times 10^9 = 3.93\times 10^{14} \text{Wei} = 3.93\times 10^{-4}~\text{ETH} \approx 0.06~\text{USD}$ to execute code for one second. Consequently, it would cost slightly more than $0.74$~USD per block to prevent nodes running on commodity hardware to keep up with the network. This is a very cheap price to pay and could indeed motivate the probable attackers discussed earlier to execute such an attack.

It is worth noting that if an attacker wanted to fill a larger portion of the block with malicious transactions, he would need to increase the gas price. Indeed, to fill half of the block with malicious transactions, it would require to pay around~\empirical{15Gwei}, or~\empirical{5} times more per gas, than to fill only~\empirical{20}\% of the block. This would result in a cost of~$10,000,000 \times 50\% \times 1.5\times 10^{10} = 0.075~\text{ETH} \approx 10.88~\text{USD}$. Nevertheless, this remains a very low price to pay for an attacker with financial incentives such as the ones described earlier.

\point{Attack limitations}
The current requirements to run a full node on the Ethereum main net are low enough for most commodity hardware to be able to keep up without any issue. The only point mentioned by the Ethereum developers is that running a full node requires an SSD~\cite{ethereum-faq}. Although there is currently no official documentation on other requirements, other sources estimate the minimum required memory to be about~8GB~\cite{node-incentive,pantheon-system-requirements,eth-hardware-requirements}. However, there is very little information about the typical hardware setup of full nodes. Therefore, it is very difficult to accurately evaluate how many nodes would be affected by such an attack. Nevertheless, the attack was judged severe enough by the Ethereum developers to react very promptly (within less than 24 hours for the first reply and within four days for them to test the fix) after our disclosure.




\subsection{Responsible Disclosure}
\label{sec:responsible}
Given that the attack is very easy and cheap to execute, and worked on all major clients, we went through a responsible disclosure process. The Ethereum Foundation has an official bug bounty program~\cite{ethereum-bug-bounty} to report vulnerabilities. With the help of colleagues\footnote{Matthias Egli and Hubert Ritzdorf from PwC Switzerland}, we wrote a report summarising our main findings, including a minimal script to execute our attack, and sent it to the bug bounty program on October~3,~2019. We received a reply the next day from the Ethereum Security Lead, acknowledging the issue and pointing us to some ongoing efforts to improve some of the inefficiencies exploited by our attack. The Ethereum foundation team also let us know that they would coordinate with Parity developers. After discussions about the ongoing efforts and some other potential solutions, we have been confirmed that our report had been awarded a reward of~5,000~USD on November~17,~2019. Finally, the official announcement was published on the bounty program website on January~7,~2020.

\section{Towards a Better Approach}
\label{sec:design}

Gas metering and pricing is a difficult but fundamental problem in Ethereum and other blockchains which use a similar approach to price contract execution. Mispricing of gas instructions has been a concern for a long time and improvements have been included in several hard forks~\cite{erc150,eip2200}. However, there remain issues in the current Ethereum pricing model, allowing attacks such as the one we presented in the previous sections. In this section, we will discuss short-term fix which can be used to prevent DoS such as the one presented in this paper, and then briefly present longer-term potential solutions which are still being actively researched. 

The main attack vector presented in this paper comes from the low speed of searching for an account which is not currently cached. One of the main issues is that the state of Ethereum gets larger with time. This means that operations accessing the state get more expensive with time in terms of resource usage.

\point{Short-term fixes}
Short-term fixes for slow IO related issues can be categorised in the two following classes: increase in the gas cost of IO instructions, as seen in EIP-150~\cite{erc150} and EIP-2200~\cite{eip2200}, and improvements in the speed of Ethereum clients.

Increasing the cost of IO instructions improves the fairness of the gas costs yet is often not sufficient to protect against DoS attacks, albeit it does increase their cost. The attack we present in this paper uses mainly instructions whose prices have increased in EIP-150 or EIP-2200, but remains relatively cheap to execute.

Improvements involve adding more layers of caches to reduce the number of IO accesses, which are typically the bottleneck. However, this requires to keep more data in memory and therefore creates a trade-off between memory consumption and execution speed. Regarding account lookup, two cases must be considered: when the looked up account exists and when it does not. Naively caching all the accounts could allow an attacker to easily evict existing accounts from the cache and is therefore dangerous. To check whether a particular account exists, a Bloom filter can be utilised as a first test. This eliminates the need for most of the IO accesses in case the queried address does not exist, while keeping a relatively low memory footprint~\cite{mitzenmacher2002compressed}. The next case which needs to be handled is the fast lookup of existing accounts. The current attempt to do this keeps an on-disk dynamic snapshot of the accounts state~\cite{dynamic-trie-snapshot-pr}, which allows to perform an on-disk look up of an account in~$\mathcal{O}(1)$, at the cost of increasing the storage usage of the node. This indeed solves the bottleneck of accessing account data but is very specific to this particular issue.

\point{Long-term fixes}
Long-term fixes are likely to only arrive in Ethereum~2.0, as most of them will require major and breaking changes. There have been several solutions discussed by the community and other researchers, which can mostly be categorised as either~a)~changing the gas pricing mechanism or~b)~changing the way clients store data.

Current proposals to change the gas mechanism involve making the pricing more dynamic that it is currently. Chen et al.~\cite{Chen2017Metering} propose a mechanism where contracts using a single instruction in excess would be penalised. The threshold is set using historical data in order to penalise only contracts which diverge too much from what would be a regular usage. Although the approach has some advantages over the current pricing mechanism, it is unclear how well it would be able to prevent attacks taking this mechanism into account.

A promising and actively researched approach is the use of stateless clients and stateless validation. The key idea is that instead of forcing clients to store the whole state, entity emitting transactions must send the transaction, the data needed by the transactions, and a proof that this data is correct. The proof can be fairly trivially constructed as a Merkle proof, as the block headers hold a hash of the root of the state and the state can be represented as a Merkle tree. This allows such clients to verify all transactions without accessing IO resources at all, making execution and storage much cheaper, at the cost of an increased complexity when creating transactions and a higher bandwidth usage.

Another active area of research which should help making things better in this direction is sharding~\cite{al2017chainspace}. Although sharding does not address the fundamental issue of gas pricing in the presence of IO operations, it does help to keep the state of the nodes smaller, as different shards will be responsible for storing the state of different parts of the network.

\section{Related Work}
\label{sec:related}

There has been a great deal of attention focused on the correctness of smart contracts on blockchains, especially, the Ethereum blockchain. Some of the vulnerability types have to do with gas consumption, but not all. There has been relatively little attention given to the organisation of metering for blockchain systems. We will first present research focusing on smart contract issues, and then highlight the work that focuses on metering at the smart contract level. We will then present research focusing on metering at the virtual machine level~---~EVM in the case of Ethereum.

\subsection{Smart Contracts}
Major contracts vulnerabilities have been observed in recent years~\cite{Atzei2017} with sometimes multiple millions of dollars worth of Ether at stake~\cite{Securities2017,parity-wallet-freeze}.
One of the most famous exploit on the Ethereum blockchain was The DAO exploit~\cite{mehar2019understanding}, where an attacker used a re-entrancy vulnerability~\cite{Luu2016a,DBLP:conf/ndss/KalraGDS18} to drain funds out of The DAO smart contract. The attacker managed to drain more than 3.5 million of Ether, which would now be worth more than 507.50 million USD. Given the severity of the attack, the Ethereum community decided to hard-fork the blockchain, preventing the attacker to benefit from the Ether he had drained.

In order to prevent such exploits, many different tools have been developed over the years to detect vulnerabilities in smart contracts~\cite{harz2018towards}. One of the first tools which have been developed is Oyente~\cite{Luu2016a}. It uses symbolic execution to explore smart contracts execution pass and then uses an SMT solver~\cite{de2008z3} to check for several classes of vulnerabilities. Many other tools covering the same or other classes of vulnerabilities have also been developed~\cite{DBLP:conf/ndss/KalraGDS18,Brent2018,Tsankov:2018:SPS:3243734.3243780,Jiang:2018:CFS:3238147.3238177} and are usually based either on symbolic execution or static analysis methods such as data flow or control flow analysis. Some smart contract analysis tools have also focused more on analysing vulnerabilities related to gas~\cite{Grech2018,Chen2017,DBLP:journals/corr/abs-1811-10403}. We present some of these tools in the next subsection.

\subsection{Gas Usage and Metering}
Recent work by Yang et al.~\cite{DBLP:journals/corr/abs-1905-00553} have recently empirically analysed the resource usage and gas usage of the EVM instructions. They provide an in-depth analysis of the time taken for each instructions both on commodity and professional hardware. Although our work was performed independently, the results we present in Section~\ref{sec:case-studies} seem to concur mostly with their findings.

Other related themes have also been covered in the literature. One of the large theme is optimisation of gas usage for smart contracts. Another one is estimating, preferably statically, the gas consumption of smart contracts.

\subsubsection*{Gas Usage Optimisation}
Gasper~\cite{Chen2017} is one of the first paper which has focused on finding gas related anti-patterns for smart contracts. It identifies 7 gas-costly patterns, such as dead code or expensive operations in loops, which could potentially be costly to the contract developer in terms of gas. Gasper builds a control flow graph from the EVM bytecode and uses symbolic execution backed by an SMT solver to explore the different paths that might be taken.

MadMax~\cite{Grech2018} is a static analysis tool to find gas-focused vulnerabilities. Its main difference with Gasper from a functionality point of view is that MadMax tries to find patterns which could cause out-of-gas exceptions and potentially lock the contract funds, rather than gas-intensive patterns. For example, it is able to detect loops iterating on an unbounded number of elements, such as the numbers of users, and which would therefore always run out of gas after a certain number of users. MadMax decompiles EVM contracts and encodes properties about them into Datalog to check for different patterns. It is performant enough to analyse all the contracts of the Ethereum blockchain in only 10 hours.

\subsubsection*{Gas Estimation}
Marescotti et al.~\cite{10.1007/978-3-030-03427-6_33} propose two algorithms to compute upper-bound gas consumption of smart contracts. It introduces a ``gas consumption path'' to encode the gas consumption of a program in its program path. It uses an SMT solver to find an environment resulting in a given path and computes its gas consumption. However, this work is not implemented with actual EVM code and is therefore not evaluated on real-world contracts.

Gastap~\cite{DBLP:journals/corr/abs-1811-10403} is a static analysis tool which allows to compute sound upper bounds for smart contracts. This ensures that if the gas limit given to the contract is higher than the computed upper-bound, the contract is assured to terminate without out-of-gas exception. It transforms the EVM bytecode and models it in terms of equations representing the gas consumption of each instructions. It then solves these equations using the equation solver PUBS~\cite{10.1007/978-3-540-69166-2_15}. Gastap is able to compute gas upper bound on almost all real world contracts it is evaluated on.

\subsection{Virtual Machines and Metering}
Zheng et al.~\cite{8449244} propose a performance analysis of several blockchain systems which leverage smart contracts. Although the analysis goes beyond smart contracts metering, with metrics such as network related performance, it includes an analysis about smart contracts metering at the virtual machine level. Notably, it shows that some instructions, such as \lstinline{DIV} and \lstinline{SDIV}, consume the same amount of gas while their consumption of CPU resource is vastly different.

Chen et al.~\cite{Chen2017Metering} propose an alternative gas cost mechanism for Ethereum. The gas cost mechanism is not meant to replace completely the current one, but rather to extend it in order to prevent DoS attacks caused by under-priced EVM instructions. The authors analyse the average number of execution of a single instruction in a contract, and model a gas cost mechanism to punish contracts which excessively execute a particular instruction. This gas mechanism allows normal contracts to almost not be affected by the price changes while mitigating spam attacks which have been seen on the Ethereum blockchain~\cite{transaction-spam-attack}.

\section{Conclusion}
\label{sec:conclusion}

In this work, we presented a new DoS attack on Ethereum by exploiting the metering mechanism. We first re-executed the Ethereum blockchain for~\Months months and showed some significant inconsistencies in the pricing of the EVM instructions. We further explored various other design weaknesses, such as gas costs for arithmetic EVM instructions and cache dependencies on the execution time. Additionally, we demonstrated that there is very little correlation between gas and resources such as CPU and memory. We found that the main reason for this is that the gas price is dominated by the amount of \emph{storage} used.

Based on our observations, we presented a new attack called~\emph{Resource Exhaustion Attack} which systematically exploits these imperfections to generate low-throughput contracts. Our genetic algorithm is able to generate programs which exhibit a throughput of around~\empirical{1.25M} gas per second after a single generation. A minimum in our experiments is attained at generation~\empirical{243} with the block using around~\empirical{9.9M} gas and taking around~\empirical{93 seconds}. We showed that we are able to generate contracts with a throughput as low as~\empirical{107,000} gas per second, or on average more than~\Slowdown times slower than typical contracts, and that all major Ethereum clients are vulnerable. We argued that several attackers such as speculators, Ethereum competitors or even miners could have financial incentives to perform such an attack. Finally, we discussed about short-term and potential long-term fixes for gas mispricing. Our attack went through the a responsible disclosure process and has been awarded a bug bounty reward of~5,000~USD by the Ethereum foundation.

\section*{Acknowledgment}
The authors would like to thank Matthias Egli and Hubert Ritzdorf from PwC Switzerland for their insightful feedback and their help with responsible disclosure and experiments.

The authors would also like to thank the Tezos Foundation for their financial support.

\bibliographystyle{plain}
\bibliography{blockchain-security}
\end{document}